\documentclass[draft]{agujournal}

\journalname{Radio Science}
\usepackage[utf8]{inputenc}
\usepackage[T1]{fontenc}
\usepackage{graphicx}
\usepackage{mathtools}
\usepackage{siunitx}
\usepackage{url}
\usepackage{soul}
\usepackage{ulem}
\usepackage{amsmath}
\usepackage{amssymb}
\usepackage{rotating}

\begin{document}

%
%

\title{Impedance and voltage power spectra of a monopole antenna in a warm plasma - derivation and application to CubeSats}

%
%




\authors{Ronald Maj\affil{1,2}, Iver H. Cairns\affil{1} and M.M. Martinovi\'c\affil{3,4,5}}


\affiliation{1}{School of Physics, The University of Sydney, NSW 2006, Australia}
\affiliation{2}{SPACE Research Centre, RMIT University, Melbourne, Victoria, Australia}
\affiliation{3}{Lunar and Planetary Laboratory, University of Arizona, Tucson, AZ 85719, USA}
\affiliation{4}{LESIA, Observatoire de Paris, Meudon, France}
\affiliation{5}{Department of Astronomy, Faculty of Mathematics, University of Belgrade, Serbia}




\correspondingauthor{Ronald Maj}{ronald.maj@rmit.edu.au}




\begin{keypoints}
\item An antenna response function is derived for monopole antennas $F_{m1}(x)$ but is only useful in certain circumstances (e.g. modelling the QTN)
\item Integration over $F_{m1}(x)$ to model the impedance of the antenna is well-behaved for the real part but does not converge for the imaginary part of the warm plasma impedance integral. The shot noise and capacitance predictions do not converge as a result. However, this is also true for the well-known double-sphere antenna response function
\item Restricting the integration on physical grounds did not lead to a useful general criterion to limit the integration. Further work is necessary to find the source of the issues with shot noise and capacitance or determine a new monopole response function
\end{keypoints}

%
%


\begin{abstract}

The impedance for a monopole antenna is derived and compared with the cases for wire dipole and double-sphere antennas. This derivation produces a new expression for the monopole antenna response function, $F_{m1}(x)$. The monopole, wire dipole, and double-sphere response functions are compared by modeling an antenna in Earth's ionospheric plasma (i.e. electrostatic and collisionless) and predicting the antenna capacitance and voltage power spectra for quasi-thermal noise (QTN) and shot noise. The monopole antenna current distribution is assumed to be a half-triangular current distribution (considering only the positive half of the triangular distribution). The predictions for the shot noise and capacitance presented problems, as the integral over wavenumber-space or $k$-space did not converge for large values of $k$. The derived expression therefore remains a current problem and necessitates future work to determine a more general expression. In this paper we bring the problem of an appropriate analytic monopole antenna response function to the attention of the community and outline a number of tests that can be used to verify any future expression. 
\end{abstract}

%
%

%


%
%
%
%

\section{Introduction}
Antennas play a vital role in communication, for example allowing signals to be sent from mobile devices to receiving towers or even between Earth-orbiting satellites and ground stations. Antennas are also important in researching the natural environment through study of the electromagnetic and electrostatic waves and other signals that may be present. To model an antenna, parameters such as the gain, radiation pattern, and impedance may need to be known. This is also true for antennas used to diagnose space plasmas through quasi-thermal noise (QTN) and shot noise spectroscopy assuming Maxwellian or Kappa velocity distribution functions \citep{couturier1981, kellogg1981, meyer1983, meyer1989, maksimovic1995, chat2009, maj2017}. Quasi-thermal noise is due to the thermal motions of plasma particles producing electrostatic Langmuir waves which can be detected with sensitive receivers. Shot noise is due to the impact of plasma particles with the antenna producing voltage peaks/troughs which can be approximated by a step function at low frequencies and are again detectable with a sensitive receiver. In these cases, the impedance of the antenna in a warm plasma needs to be known to predict the power spectrum and therefore determine the plasma properties. 

Authors such as \citet{balmain1965}, \citet{kuehl1966}, \citet{meyer1979}, and \citet{couturier1981} have outlined derivations for the impedance of a dipole antenna assuming a triangular current distribution. However, the literature does not appear to contain a derivation for a monopole antenna. We derive the monopole impedance following the general method outlined in \citet{kuehl1966} and \citet{couturier1981}, expanding on intermediate steps and using a different expression for the current distribution, namely one for a monopole antenna. Through this process a new expression for the antenna response function is derived and compared to the wire dipole response function $F_1(x)$ introduced in \citet{kuehl1966} as well as to the response function for a double-sphere antenna \citep{meyer1989}. This response function is the main difference between the monopole, wire dipole and double-sphere calculations of the impedance. 

This paper proceeds by presenting the theoretical background for the impedance in Section \ref{sec:Gen_imped}. The derivations for the dipole and monopole impedance are then detailed in Sections \ref{sec:Der_Di} and \ref{sec:Der_Mono}, respectively. The ensuing predictions for the voltage power spectrum due to quasi-thermal noise (QTN) and shot noise are outlined in Section \ref{sec:Compare}.  The capacitances for dipole and monopole antennas calculated using the derived impedances, plus their comparisons with standard expressions in the low frequency approximation, are presented in Section \ref{sec:Compare}. These results are discussed and the possibility of restricting the maximum wavenumber $k$ value based on physical grounds is investigated in Section \ref{sec:Discuss}. Finally, Section \ref{sec:Conc} concludes the paper.

\section{General impedance expression}
\label{sec:Gen_imped}
Determining the impedance of an antenna in a plasma requires a model representation of the plasma. The most common ways of doing this involve using either the magnetohydrodynamic equations (which are well suited to describe the plasma macroscopically) or the Vlasov equations (which take a kinetic approach and reveal microscopic details). The derivation here will use the kinetic approach and follow the work of \citet{kuehl1966} and \citet{couturier1981}. In this section we derive the resistance of a wire dipole antenna of radius $a$ and length $L$ for each arm. Firstly, we will assume the following \citep{kuehl1966}:
\begin{itemize}
\item the plasma is unmagnetized
\item the plasma is at thermal equilibrium (Maxwellian and non-zero temperature, i.e. warm),
\item collisions are neglected,
\item ion motions are neglected,
\item the effects of electron and ion sheaths around the antenna are neglected \citep{meyer1993},
\item the field is weak enough to allow linearized equations (any external electromagnetic field is also neglected),
\item the antenna is long with length $L\gg \lambda_D$ where $\lambda_D = \sqrt{\epsilon_0 k_B T_e / e^2 n_e } $ is the electron Debye length (electron density and temperature $n_e$ and $T_e$, respectively), $\epsilon_0$ the permittivity of free space, $k_B$ the Boltzmann constant, and $e$ the electric charge, and
\item the antenna radius $a$ is finite with $a\ll L$ and $a\ll \lambda_D$.
\end{itemize}
Following \citet{couturier1981} the antenna impedance $Z_a$ is given by  
\begin{linenomath*}
\begin{equation}
\label{eq:Imped}
Z_a = R - iX = -\frac{1}{I_0^2}\int \vec{E}(\vec{r})\cdot\vec{J}(\vec{r}) \hspace{3pt} d\vec{r}
\end{equation}
\end{linenomath*}
where $R$ is the resistance, $X$ the reactance, $i$ the imaginary unit, $I_0$ is the peak current flowing into/out of the antenna, $\vec{E}(\vec{r})$ the electric field of the source, and $\vec{J}(\vec{r})$ the current distribution. Fourier transforming and using Parseval's theorem allows us to write \citep{couturier1981} 
\begin{linenomath*}
\begin{equation}
\label{eq:Imped_k}
Z_a = \frac{i}{I_0^2(2\pi)^3\omega\epsilon_0}\int J_i^*(\vec{k}) \Lambda_{ij}^{-1}(\vec{k},\omega)J_j(\vec{k}) \hspace{3pt} d\vec{k}
\end{equation}
\end{linenomath*}
where the summation over dummy indices is implied, $J^*_i$ is the complex conjugate of $J_i$ and $\Lambda_{ij}^{-1}(\vec{k},\omega)$ is defined with respect to the plasma dielectric permittivity tensor $\epsilon_{ij}(\vec{k},\omega)$ as \citep{sitenko1967}
\begin{linenomath*}
\begin{equation}
\label{eq:Lambdaij}
\Lambda_{ij}(\vec{k},\omega) =\frac{k^2c^2}{\omega^2} \left(\frac{k_ik_j}{k^2} - \delta_{ij}\right) + \epsilon_{ij}(\vec{k},\omega).
\end{equation}
\end{linenomath*}
To simplify, we consider only the longitudinal component and so (\ref{eq:Imped_k}) yields \citep{meyer1989}
\begin{linenomath*}
\begin{equation}
\label{eq:Imped_L}
Z_a = \frac{i}{I_0^2(2\pi)^3\omega\epsilon_0}\int \frac{|\vec{k}\cdot\vec{J}(\vec{k})|^2}{k^2\epsilon_L}\hspace{3pt} d\vec{k}
\end{equation}
\end{linenomath*}
as the impedance for the antenna. The current distribution $\vec{J}(\vec{k})$ used will determine the final form that (\ref{eq:Imped_L}) takes. The longitudinal dielectric permittivity $\epsilon_L$ is defined as
\begin{linenomath*}
\begin{equation}
\label{eq:epsL}
\epsilon_L= 
1 + \frac{1}{k^2\lambda_D^2} +\frac{2i\omega\sqrt{\pi} W\left(\frac{\omega}{k v_T}\right)}{k^2\lambda_D^2} \left(\frac{\omega_p}{k v_T}\right)^2
\end{equation}
\end{linenomath*}
where $\omega_p = 2 \pi f_p$ is the angular electron plasma frequency, $v_T = \sqrt{\frac{2k_BT}{m_e}}$ is the thermal speed of electrons with mass $m_e$, $W(z)$ is the Faddeeva function defined as
\begin{linenomath*}
\begin{equation}
\label{eq:Fadd}
W(z) = e^{-z^2} \mathrm{erfc}(-iz),
\end{equation}
\end{linenomath*}
and $\mathrm{erfc}(z)$ is the complementary error function. 

The dielectric permittivity tenor $\epsilon_{ij}$ defines how waves will travel in the given plasma environment. We have decided to use the Earth's ionospheric environment at low latitudes and low altitudes ($\approx$ \SI{300}{\kilo\meter}) as our test case to derive the expressions for the antenna response functions for a dipole and monopole antenna. This is to say that the plasma is assumed to be at thermal equilibrium and unmagnetized, allowing the dielectric permittivity tensor to be separated into independent longitudinal and transverse components, which is why we only consider $\epsilon_L$ in our equations. 

The unmagnetized condition is not entirely accurate as the geomagnetic field strength is on the order of \SIrange{20}{60}{\micro\tesla} at \SI{300}{\kilo\meter} altitude, according to the International Geomagnetic Reference Field model \citep{thebault2015}. This places the electron cyclotron frequency between \SI{0.6}{\mega\hertz} and \SI{1.7}{\mega\hertz}. Using the International Reference Ionosphere (IRI) model data for electron density and temperature, the plasma frequency $f_p$ ranges from \SIrange{1.5}{8.9}{\mega\hertz} at \SI{300}{\kilo\meter} altitude. Therefore the effects of the magnetic field are significant enough to cause appreciable effects to the plasma waves predicted in a voltage power spectrum, for instance shifting the Langmuir wave frequency away from $\omega_p $ and having additional thermal Bernstein and upper hybrid waves modes detectable. However, in this paper we will concentrate on the unmagnetized case as a first order approximation. Also, these results could be extended down to altitudes of \SIrange{100}{140}{\kilo\meter} where the plasma is mostly unmagnetized and approximately collisionless (although the effect of collisions are notable at around \SI{120}{\kilo\meter} \citep{martinovic2017}).

Relaxing the thermal equilibrium condition requires a different expression for the velocity distribution function (VDF) of the plasma particles. In this paper we will compare the predictions for the Maxwellian VDF against those for a Kappa VDF in order to explore any differences between thermal and non-thermal conditions. In effect this involves substituting the longitudinal dielectric permittivity $\epsilon_L$ in (\ref{eq:epsL}) with \citep{chateau1991}
\begin{linenomath*}
\begin{equation}
    \label{eq:epsL_kap}
    \epsilon_L = 1 + \frac{z^2}{r^2}\left( 2\kappa -1 + \frac{(-2)^{\kappa + 1}}{(2\kappa - 3)!!}iz \sum_{p=0}^\kappa \frac{(\kappa + p)!}{p!}\frac{1}{(2i)^{\kappa+1+p}(z+i)^{\kappa+1-p}} \right) 
\end{equation}
\end{linenomath*}
where $z = \omega /(k v_0\sqrt{\kappa})$, $r = \omega / \omega_p$, and $v_0 = v_T\sqrt{(2\kappa-3)/2\kappa}$. It should be noted that the formalism in \citet{chateau1991} and expressed above is for integer values of $\kappa$ only, and a more general expression can be found in \citep{chat2009}. Using a Kappa distribution also alters the definition of the Debye length as the value of $\kappa$ affects the distance over which plasma particles are shielded. We will call this modified Debye length $\lambda_{D-\kappa}$ with \citep{chateau1991} 
\begin{linenomath*}
\begin{equation}
    \label{eq:LD_kap}
    \lambda_{D-\kappa} = \frac{v_0}{\omega_p}\sqrt{\frac{\kappa}{2\kappa - 1}}.
\end{equation}
\end{linenomath*}

Other distributions such as a Flat-Top or Heaviside VDF could also be used to reveal non-thermal effects. Kappa distributions with values of $\kappa \approx 2 - 4$ for the Earth's magnetosphere \citep{vasyliunas1968} and $\kappa \approx 4 - 7$ in the solar wind \citep{chateau1991,maksimovic1997a,maksimovic1997} provide a good fit to measured energy spectra and other observations. Therefore we will use $\kappa = 4$ for our predictions in this paper. 

In Section \ref{sec:Compare} we look at the voltage power spectra predicted for QTN and shot noise as well as the predicted capacitance for this environment. Based on the assumption of thermal equilibrium, the expression for QTN has the form \citep{meyer1989}
\begin{linenomath*}
\begin{equation}
\label{eq:QTN_gen}
V_{QTN}^2 = 4k_BT_e \Re(Z_a)
\end{equation}
\end{linenomath*}
where $\Re(Z_a)$ denotes the real part of the impedance $Z_a$, that is, the resistance of the antenna. This can be expressed as $R = R_L + R_T$, which are the longitudinal ($L$) and transverse ($T$) components of the resistance, which correspond to electrostatic and electromagnetic waves, respectively. In the non-thermal Kappa-distribution case, the QTN expression we will use is \citep{chateau1991}
\begin{linenomath*}
\begin{equation}
\label{eq:QTN_kap}
V_{QTN-\kappa}^2 = \frac{2^{\kappa +3}}{\pi^2\epsilon_0} \frac{\kappa !}{(2\kappa -3)!!} \frac{mv_0}{r^2} \int_0^{+\infty} z F\left( \frac{ru}{z\sqrt{2\kappa - 1}}\right)[(1+z)^\kappa|\epsilon_L|^2]^{-1} dz
\end{equation} 
\end{linenomath*}
where $u=L/\lambda_{D-\kappa}$ and the integral is over $z$ as defined previously. The shot noise has the form \citep{meyer1989}
\begin{linenomath*}
\begin{equation}
\label{eq:Shot_gen}
V_{S}^2 = 2e^2 N_e |Z_a|^2
\end{equation}
\end{linenomath*}
where $|Z_a|^2 = Z_a^* \times Z_a$ with $Z_a^*$ the complex conjugate of $Z_a$, and $N_e = 1/{\sqrt{4\pi}}n_e v_T S$ is the impact rate of plasma electrons
with $S$ the antenna (or satellite) surface area. Equation (\ref{eq:Shot_gen}) is only valid below the electron plasma frequency \citep{meyer1983}. 

For the capacitance we use
\begin{linenomath*}
\begin{equation}
\label{eq:Capacit_imped}
C_a = \frac{1}{\omega\hspace{3pt} \mathrm{Im}(Z_a)}
\end{equation}
\end{linenomath*}
which is the case for an ideal capacitor and involves the antenna impedance $Z_a$ for an arbitrary antenna. We can also calculate the capacitance by using the analytic approximation for the reactance \citep{meyer1989} of a long dipole ($L \gg \lambda_D$), which gives
\begin{linenomath*}
\begin{equation}
\label{eq:Capacit_dipole}
C_a = \frac{\pi\epsilon_0 L}{\log{(\lambda_D/a)}}
\end{equation}
\end{linenomath*}
for a wire dipole and 
\begin{linenomath*}
\begin{equation}
\label{eq:Capacit_dipole_sph}
C_a = 2\pi\epsilon_0a
\end{equation}
\end{linenomath*}
for a double-sphere antenna.

The expressions (\ref{eq:Capacit_dipole}) and (\ref{eq:Capacit_dipole_sph}) are only valid for low frequencies $f\ll f_p$ . Therefore in the dipole case, we are able to compare the approximations (\ref{eq:Capacit_dipole}) and (\ref{eq:Capacit_dipole_sph}) with the capacitance calculated from the dipole impedance and (\ref{eq:Capacit_imped}). For a monopole placed a small distance perpendicularly from a reference plane we expect that $C_{monopole} = 2C_{dipole}$. In this case the monopole creates an image resembling the dipole configuration but with only one half the voltage input/output and therefore double the capacitance \citep{balanis2016}.    

The impedance is therefore a critical part of both the QTN and shot noise expressions. The expression for the impedance that we use in this paper is based on Equation (15) in \citet{meyer1989}. Specifically we will aim to derive an expression of the form
\begin{linenomath*}
\begin{equation}
\label{eq:Imped_w/F}
Z_a = \frac{4i}{\pi^2\omega\epsilon_0} \int_0^{\infty} \frac{F(k)}{\epsilon_L}\hspace{3pt}dk
\end{equation}
\end{linenomath*}
where $F(x)$ is the appropriate antenna response function. Depending on the type of antenna and the current distribution over the antenna, the form of $F(x)$ will vary. For a wire dipole $F(k)$ is given by \citep{meyer1989}
\begin{linenomath*}
\begin{equation}
\label{eq:Fk}
F(k) = F_1(kL) J_0^2(ka)
\end{equation}
\end{linenomath*}
where $J_0^2(x)$ is the zeroth Bessel function of the first kind and $F_1(x)$ takes the form
\begin{linenomath*}
\begin{equation}
\label{eq:F1}
F_1(x) = \frac{x\left(\mathrm{Si}(x) - \frac{1}{2}\mathrm{Si}(2x)\right) - 2\sin^4\left(\frac{x}{2}\right)}{x^2}
\end{equation}
\end{linenomath*}
where $\mathrm{Si}(x) = \int_0^x \frac{\sin(t)}{t} dt$ is the sine integral. Equation (\ref{eq:Fk}) assumes that $a$ is finite and in the case of an infinitely thin antenna (\ref{eq:Fk}) simplifies to $F(k) = F_1(kL)$.

The function $F_1(x)$ was derived by \citet{kuehl1966}, and has been used by many authors since then, including \citet{couturier1981}, \citet{kellogg1981}, \citet{meyer1989}, \citet{chat2009}, \citet{martinovic2016} and \citet{maj2017}. Section \ref{sec:Der_Di} derives the expression (\ref{eq:F1}), while in Section \ref{sec:Der_Mono} two new expressions are derived for the monopole response function to be used as $F(x)$ in (\ref{eq:Imped_w/F}).  

\section{Derivation of the Dipole Response Function}
\label{sec:Der_Di}

To derive the dipole response function, we begin by assuming a triangular current distribution along the antenna \citep{balmain1965}. This is expressed as
\begin{linenomath*}
\begin{equation}
\label{eq:j_r}
\vec{J}(\vec{r})=
\begin{cases}
    \frac{I_0}{2\pi a}\left(1 - \frac{|z|}{L} \right)\delta({r} - a)\hspace{3pt}\hat{z},& \text{for } |z|<L\\
    0.              & \text{for } |z|>L
\end{cases} 
\end{equation}
\end{linenomath*}
Here cylindrical coordinates are used, so that the delta function means that all the current is on surface of the antenna (${r} = a$), the antenna is aligned along the $z$-axis, and $I_0$ is the peak current flowing into/out of the antenna.

If a function is separable, i.e. $f(x_1,x_2,...) = f(x_1)f(x_2)...$, then Fourier transforming allows us to write $\mathcal{F}[f(x_1,x_2,...)] = \mathcal{F}[f(x_1)]\mathcal{F}[f(x_2)]...$ where $\mathcal{F}$ represents the Fourier transformation operation
\begin{linenomath*}
\begin{equation}
\label{eq:Fourier}
\mathcal{F}[f(x)] = \int_{-\infty}^\infty f(x) e^{-i\vec{k}\cdot\vec{r}}\hspace{5pt} d\vec{r}.
\end{equation}
\end{linenomath*}
Taking the Fourier transform of the current distribution (\ref{eq:j_r}) gives
\begin{linenomath*}
\begin{equation}
\begin{split}
\label{eq:j_k}
\mathcal{F}[\vec{J}(\vec{r})] = \vec{J}(\vec{k}) &=  I_0 J_0(k_\perp a) \int_{-\infty}^\infty \left( 1 - \frac{|z|}{L} \right) e^{-i{k_z}z}\hspace{3pt} dz \hspace{3pt}\hat{z}\\
&= I_0 J_0(k_\perp a) \left[ \int_{-L}^0 \left(1+\frac{z}{L}\right)e^{-i{k_z}z}\hspace{3pt} dz + \int_0^L \left(1-\frac{z}{L}\right)e^{-i{k_z}z}\hspace{3pt} dz\right]\hspace{3pt}\hat{z}\\
&= I_0 J_0(k_\perp a) \left[ \int_{-L}^L e^{-i{k_z}z}\hspace{3pt} dz + \int_{-L}^0 \frac{ze^{-i{k_z}z}}{L}\hspace{3pt} dz - \int_0^L \frac{ze^{-i{k_z}z}}{L}\hspace{3pt} dz \right]\hspace{3pt}\hat{z}
\end{split}
\end{equation}
\end{linenomath*}
Integration by parts $\left(\int \mathrm{u}\mathrm{v}' = [\mathrm{u}\mathrm{v}] - \int \mathrm{u}'\mathrm{v} \right)$ can be used to solve the second and third integrals above with $\mathrm{u} = z$ and $\mathrm{v}' = e^{-ik_zz}$. This then gives
\begin{linenomath*}
\begin{equation}
\begin{split}
\label{eq:j_k2}
\vec{J}(\vec{k}) &= I_0 J_0(k_\perp a) \left[ \frac{ie^{-i{k_z}L} - ie^{i{k_z}L}}{k_z} + \frac{1}{k_z L}\left(iLe^{i{k_z}L} - iLe^{-i{k_z}L} - \frac{(e^{i{k_z}L}+e^{-i{k_z}L}-2)}{k_z}\right) \right]\hspace{3pt}\hat{z}\\
&= 2 I_0 J_0(k_\perp a) \left[\frac{1 - \cos({k_z}L)}{k_z^2L} \right]\hspace{3pt}\hat{z}\\
&= \frac{4I_0^2}{k_z^2L}\sin^2\left(\frac{k_zL}{2}\right)J_0(k_\perp a)\hspace{3pt}\hat{z}
\end{split}
\end{equation}
\end{linenomath*}
using the trigonometric identity $\cos(2\theta) = 1 - 2\sin^2(\theta)$. 

The expression for the current distribution $\vec{J}(\vec{k})$ in (\ref{eq:j_k2}) can now be substituted into (\ref{eq:Imped_L}) to give
\begin{linenomath*}
\begin{equation}
\label{eq:Imped_Jsub}
Z_a = \frac{16i}{(2\pi)^3\omega\epsilon_0}\int\frac{\sin^4\left(\frac{k_zL}{2}\right)J_0^2(k_\perp a)}{k^2k_z^2L^2\epsilon_L}\hspace{3pt} d\vec{k}
\end{equation}
\end{linenomath*}
which if we use spherical coordinates in $\vec{k}$-space and carry out the $\phi$ integration can be simplified to
\begin{linenomath*}
\begin{equation}
\begin{split}
\label{eq:Imped_phi}
Z_a &= \frac{16i}{(2\pi)^3\omega\epsilon_0} \int_0^{2\pi}\int_0^{\pi}\int_0^{\infty} \frac{\sin^4\left(\frac{k_zL}{2}\right)J_0^2(k_\perp a)}{k^2k_z^2L^2\epsilon_L}\hspace{3pt} k^2 \sin(\theta) dk d\theta d\phi\\
&= \frac{4i}{\pi^2\omega\epsilon_0} \int_0^{\pi}\int_0^{\infty} \frac{\sin^4\left(\frac{k\cos(\theta)L}{2}\right)J_0^2(k\sin(\theta) a)}{k^2\cos^2(\theta)L^2\epsilon_L}\hspace{3pt} \sin(\theta) dk d\theta\\
\end{split}
\end{equation}
\end{linenomath*}
where the $\theta$ dependence has been made explicit in the second line. As we have assumed that $a\ll L$ we can approximate $J_0(k_\perp a)$ with unity since $k_\perp a \ll 1$ \citep{couturier1981}. Concentrating only on the $\theta$ integral and making the substitution $u=\cos(\theta)$ gives
\begin{linenomath*}
\begin{equation}
\begin{split}
\label{eq:Theta_int}
\int_0^{\pi} \frac{\sin^4\left(\frac{k\cos(\theta)L}{2}\right)}{k^2\cos^2(\theta)L^2}\hspace{3pt} \sin(\theta) d\theta &= \int_{-1}^1 \frac{\sin^4\left(\frac{kLu}{2}\right)}{k^2L^2u^2}\hspace{3pt} du. 
\end{split}
\end{equation}
\end{linenomath*}
Note that as the plasma is isotropic $\epsilon_L$ has only radial $k$ dependence, remaining constant over the $\theta$ integration. Using integration by parts with $\mathrm{u} = \sin^4\left({kLu}/{2}\right)$ and $\mathrm{v}' = 1/k^2L^2u^2$ in (\ref{eq:Theta_int}) gives
\begin{linenomath*}
\begin{equation}
\begin{split}
\label{eq:Theta_int_parts}
\int_{-1}^1 \frac{\sin^4\left(\frac{kLu}{2}\right)}{k^2L^2u^2}\hspace{3pt} du &= \left[ \frac{-\sin^4\left(\frac{kLu}{2}\right)}{k^2L^2u} \right]_{-1}^1 + \int_{-1}^1 \frac{1}{4kLu} (2\sin(kLu) - \sin(2kLu))\hspace{3pt}du \\
&= \frac{-2\sin^4\left(\frac{kL}{2}\right)}{k^2L^2} + \int_{0}^1\frac{\sin(kLu)}{kLu}\hspace{3pt}du - \int_{0}^1\frac{\sin(2kLu)}{2kLu}\hspace{3pt}du\\ 
&= \frac{\mathrm{Si}(kLu)}{kLu} - \frac{\mathrm{Si}(2kLu)}{2kLu} - \frac{2\sin^4\left(\frac{kL}{2}\right)}{k^2L^2}\\
&= F_1(kL)\\
\end{split}
\end{equation}
\end{linenomath*}
where $F_1(x)$, with $x = kL$, is the dipole antenna response function defined in (\ref{eq:F1}) and we have used the fact that integrals over even functions have the property that $\int_{-a}^af(x)dx = 2\int_{0}^af(x)dx$. Placing this back into (\ref{eq:Imped_phi}) gives
\begin{linenomath*}
\begin{equation}
\label{eq:Imped_w/F1}
Z_a = \frac{4i}{\pi^2\omega\epsilon_0} \int_0^{\infty} \frac{F_1(kL)J_0^2(ka)}{\epsilon_L}\hspace{3pt}dk,
\end{equation}
\end{linenomath*}
which is Equation (15) in \citet{meyer1989} or (\ref{eq:Imped_w/F}) in this text with $F_1(x)$ in place of $F(x)$. 

In summary, we have shown in this section how the antenna response function in (\ref{eq:F1}) and the  antenna impedance expression in (\ref{eq:Imped_w/F1}) are derived for the dipole case. We will use a similar procedure in the next section to derive a monopole antenna response function $F_{m1}(x)$ in order to retain the general expression (\ref{eq:Imped_w/F}) as the antenna impedance, with $F_1(x)$ replaced by the monopole function $F_{m1}(x)$.

\section{Derivation of the Monopole Response Function}
\label{sec:Der_Mono}

To derive the monopole response function, we need to assume a certain current distribution across the antenna. A monopole antenna fed with current from one end will have a peak of $I_0$ at one end and zero current at the other, so we choose a half-triangular current distribution. However the end points of the antenna can be arbitrarily placed on the axis along which the antenna lies. In this paper, we define the end points of the antenna to be at $z = -L/2$ and $z=L/2$; however, the method below produces the same results using $z=0$ and $z=L$. The current distribution is defined as
\begin{linenomath*}
\begin{equation}
\label{eq:j_r_mon1}
\vec{J}(\vec{r})=
\begin{cases}
    \frac{I_0}{2\pi a}\left(\frac{1}{2} - \frac{z}{L} \right)\delta(\vec{r} - a)\hspace{3pt}\hat{z},& \text{for } |z|<\frac{L}{2}\\
    0,              & \text{for } |z|>\frac{L}{2}
\end{cases} 
\end{equation}
\end{linenomath*}
for an antenna aligned along the $z$-axis. Taking the Fourier transform of (\ref{eq:j_r_mon1}) gives
\begin{linenomath*}
\begin{equation}
\begin{split}
\label{eq:j_k_mon1}
\vec{J}(\vec{k}) &=  I_0 J_0(k_\perp a) \int_{-L/2}^{L/2} \left( \frac{1}{2} - \frac{z}{L} \right) e^{-i{k_z}z}\hspace{3pt} dz \hspace{3pt}\hat{z}\\
&= I_0 J_0(k_\perp a) \left[ \frac{1}{2}\int_{-L/2}^{L/2}e^{-i{k_z}z}\hspace{3pt} dz -\frac{1}{L} \int_{-L/2}^{L/2} ze^{-i{k_z}z}\hspace{3pt} dz\right]\hspace{3pt}\hat{z}.
\end{split}
\end{equation}
\end{linenomath*}
Carrying out the $z$ integration (\ref{eq:j_k_mon1}) becomes
\begin{linenomath*}
\begin{equation}
\begin{split}
\label{eq:j_k2_2}
\vec{J}(\vec{k}) &= I_0 J_0(k_\perp a) \left[ \frac{ie^{-i{k_z}L/2} - ie^{i{k_z}L/2}}{2k_z} - \frac{1}{k_z L}\left( \frac{iL}{2}(e^{-i{k_z}L/2} - e^{i{k_z}L/2}) - \frac{(e^{i{k_z}L/2}-e^{-i{k_z}L/2})}{k_z}\right) \right]\hspace{3pt}\hat{z}\\
&= I_0 J_0(k_\perp a) \left[ \frac{2i\sin\left(\frac{k_zL}{2}\right)}{k_z^2 L}  - \frac{ie^{ik_zL/2}}{k_z} \right]\hspace{3pt}\hat{z}\\
\end{split}
\end{equation}
\end{linenomath*}
where integration by parts is used for the $ze^{-i{k_z}z}$ integral (with $\mathrm{u} = z$ and $\mathrm{v}' = e^{-ik_zz}$).

Now calculating $|\vec{k}\cdot\vec{J}(\vec{k})|^2$ for use in (\ref{eq:Imped_L}) we obtain
\begin{linenomath*}
\begin{equation}
\begin{split}
\label{eq:|kdotJ|_mon1}
|\vec{k}\cdot\vec{J}(\vec{k})|^2 &= I_0^2 k_z^2 J_0^2(k_\perp a)\left(\frac{2i\sin\left(\frac{k_zL}{2}\right)}{k_z^2 L}  - \frac{ie^{ik_zL/2}}{k_z} \right)\left( \frac{-2i\sin\left(\frac{k_zL}{2}\right)}{k_z^2 L}  + \frac{ie^{-ik_zL/2}}{k_z} \right)\\
&= I_0^2 J_0^2(k_\perp a) \left(\frac{4\sin^2\left(\frac{k_zL}{2}\right)}{k_z^2 L^2}  - \frac{2\sin(k_zL)}{k_zL} + 1 \right).  \\
\end{split}
\end{equation}
\end{linenomath*}
After substituting (\ref{eq:|kdotJ|_mon1}) into (\ref{eq:Imped_L}), using spherical coordinates, and carrying out the $\phi$ integration we obtain
\begin{linenomath*}
\begin{equation}
\begin{split}
\label{eq:Imped_Mon1}
Z_a = \frac{i}{4\pi^2\omega\epsilon_0}\int_0^\pi\int_0^\infty \left(\frac{4\sin^2\left(\frac{kL\cos(\theta)}{2}\right)}{k^2 L^2 \cos^2(\theta)}  - \frac{2\sin(kL\cos(\theta))}{kL\cos(\theta)} + 1 \right)\frac{J_0^2(ka\sin(\theta))\sin(\theta)}{\epsilon_L}\hspace{3pt} d{k}d\theta
\end{split}
\end{equation}
\end{linenomath*}
which can be simplified to create a new antenna response function $F_{m1}(x)$ by concentrating on the $\theta$ integral. Using the substitution $u=\cos(\theta)$ we write the $\theta$ integral as
\begin{linenomath*}
\begin{equation}
\begin{split}
\label{eq:Theta_Int_mon1}
&\int_0^\pi\left(\frac{4\sin^2\left(\frac{kL\cos(\theta)}{2}\right)}{k^2 L^2 \cos^2(\theta)}  - \frac{2\sin(kL\cos(\theta))}{kL\cos(\theta)} + 1 \right)\sin(\theta)\hspace{3pt} d\theta \\
&= \int_{-1}^1 \left(\frac{4\sin^2\left(\frac{kLu}{2}\right)}{k^2 L^2 u^2}\right)\hspace{3pt}d{u} -2\int_{-1}^1\frac{\sin(kLu)}{kLu}\hspace{3pt}d{u} + \int_{-1}^1 1\hspace{3pt}d{u}\\
&= \left[ \frac{-4\sin^2\left(\frac{kLu}{2}\right)}{k^2L^2u} \right]_{-1}^1 + \int_{-1}^1 \frac{4}{kLu}\sin\left(\frac{kLu}{2}\right)\cos\left(\frac{kLu}{2}\right)\hspace{3pt}d{u} - \frac{4\mathrm{Si}(kL)}{kL} + 2\\
&= 2 - \frac{8\sin^2\left(\frac{kL}{2}\right)}{k^2L^2} - \frac{4\mathrm{Si}(kL)}{kL} + 2\int_{-1}^1\frac{\sin(kLu)}{kLu}\hspace{3pt}d{u}\\
&= 2\left(\frac{k^2L^2 - 4\sin^2\left(\frac{kL}{2}\right)}{k^2L^2} \right)
\end{split}
\end{equation}
\end{linenomath*}
where integration by parts was used in the first integral on the second line with $\mathrm{u} = \sin^2\left({kLu/2}\right)$ and $\mathrm{v}' = {4/k^2L^2u^2}$. Placing (\ref{eq:Theta_Int_mon1}) back into (\ref{eq:Imped_Mon1}) yields
\begin{linenomath*}
\begin{equation}
\begin{split}
\label{eq:Imped_Mon1_sub}
Z_a = \frac{i}{2\pi^2\omega\epsilon_0}\int_0^\infty \left(\frac{k^2L^2 - 4\sin^2\left(\frac{kL}{2}\right)}{k^2L^2} \right)\frac{J_0^2(ka)}{\epsilon_L}\hspace{3pt} d{k}.
\end{split}
\end{equation}
\end{linenomath*}
We now define $F_{m1}(x)$, the monopole response function, as 
\begin{linenomath*}
\begin{equation}
\label{eq:F3}
F_{m1}(x) = \frac{1}{8} \left( 1 - \frac{\sin^2\left(\frac{x}{2}\right)}{2x^2} \right) 
\end{equation}
\end{linenomath*}
where $x = kL$ and the constant $1/8$ is introduced to match the form of (\ref{eq:Imped_w/F1}), so that (\ref{eq:Imped_Mon1_sub}) now becomes
\begin{linenomath*}
\begin{equation}
\begin{split}
\label{eq:Imped_Mon1_w/F3}
Z_a = \frac{4i}{\pi^2\omega\epsilon_0}\int_0^\infty\frac{F_{m1}(kL)J_0^2(ka)}{\epsilon_L}\hspace{3pt} d{k}.
\end{split}
\end{equation}
\end{linenomath*}
This can now be used in the same way as (\ref{eq:Imped_w/F1}) to determine the resistance and capacitance of the antenna, and the spectra for QTN and shot noise. For large values of $x$, (\ref{eq:F3}) converges towards $1/8$ while for small values, the leading terms of the Maclaurin series expansion give $1/8\times(7/8 + x^2/96)$. 

The result for the monopole antenna response function (\ref{eq:F3}) is very similar to that of a spherical (rather than wire) dipole antenna. The antenna response function for a double-sphere dipole antenna is \citep{meyer1989, chat2009}
\begin{linenomath*}
\begin{equation}
\label{eq:F_sph}
F_{s1}(x) = \frac{1}{4}\left( 1 - \frac{\sin(x)}{x} \right)
\end{equation}
\end{linenomath*}
which in the limit of large $x$ converges to $1/4$ and the Maclaurin series has a leading term of $x^2 / 24$. For a double-sphere dipole of finite $a$, the $F(k)$ function in (\ref{eq:Imped_w/F}) would be 
\begin{linenomath*}
\begin{equation}
\label{eq:F_F_sph_sin}
F(k) = F_{s1}(kL) \frac{\sin(ka)}{k^2a^2}.
\end{equation}
\end{linenomath*}
In the case of an infinitely small $a$, this simplifies to $F(k) = F_{s1}(kL)$.

\section{Comparisons for Dipole versus Monopole Antennas}
\label{sec:Compare}
Here we compare the three antenna response functions $F_1(x)$ and $F_{s1}(x)$ for dipoles and $F_{m1}(x)$ for a monopole and the effects they have on generic spectra. We derived $F_1(x)$ and $F_{m1}(x)$ in (\ref{eq:F1}) and (\ref{eq:F3}), respectively, and showed the final expression for $F_{s1}$ in (\ref{eq:F_sph}). We compare the functions in isolation as a function of $x$ first and then within expressions that represent antennas immersed in an isotropic, unmagnetized thermal (or non-thermal) plasma like the ionosphere. The second step allows us to investigate the voltage power spectrum for QTN and shot noise, as well as the predicted capacitance and resistance of the antennas.

\begin{figure}
\centering
\includegraphics[width=0.55\linewidth]{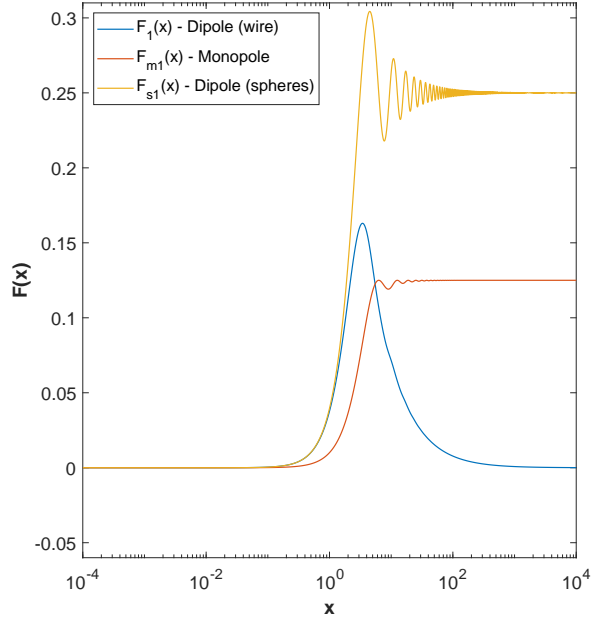}
\caption{Comparison of the antenna response function for a wire dipole antenna $F_1(x)$, as in Equation (17) of \citet{kuehl1966} or (\ref{eq:F1}) in this text, a spherical dipole antenna $F_{s1}$ using (\ref{eq:F_sph}), and the monopole expression $F_{m1}(x)$ in (\ref{eq:F3}).} 
\label{fig:Fn_vs_x}
\end{figure}

Figure \ref{fig:Fn_vs_x} compares the wire and double-sphere dipole response functions $F_1(x)$ and $F_{s1}(x)$, with the monopole response function $F_{m1}(x)$. The wire dipole function has a Gaussian-like form, peaks near $x=3.5$ and converges to zero as $x\rightarrow0$ and $x\rightarrow\infty$. The monopole function $F_{m1}(x)$ is approximately zero for small $x$ but begins to rise near $x \gtrsim 1$, peaks at 0.125 at $x \approx 6.5$, and then has an oscillatory form. At larger $x$ this oscillation dampens and $F_{m1}(x) \rightarrow 0.125$ as $x\rightarrow \infty$. The double-sphere dipole function has a similar rising and then oscillatory behavior albeit with $F_{s1}(x) \rightarrow 0.25$ as $x\rightarrow \infty$ instead. However, the peak of the function significantly overshoots its asymptotic value and dampens while it oscillates to the value 0.25. Also, the peak value is reached sooner for the two dipole response functions compared to the monopole function. Overall, the similar functional form means the monopole response function is expected to produce similar results to the double-sphere dipole case for QTN and shot noise. This will be seen in the figures that follow.

\begin{figure}
\includegraphics[width=\textwidth]{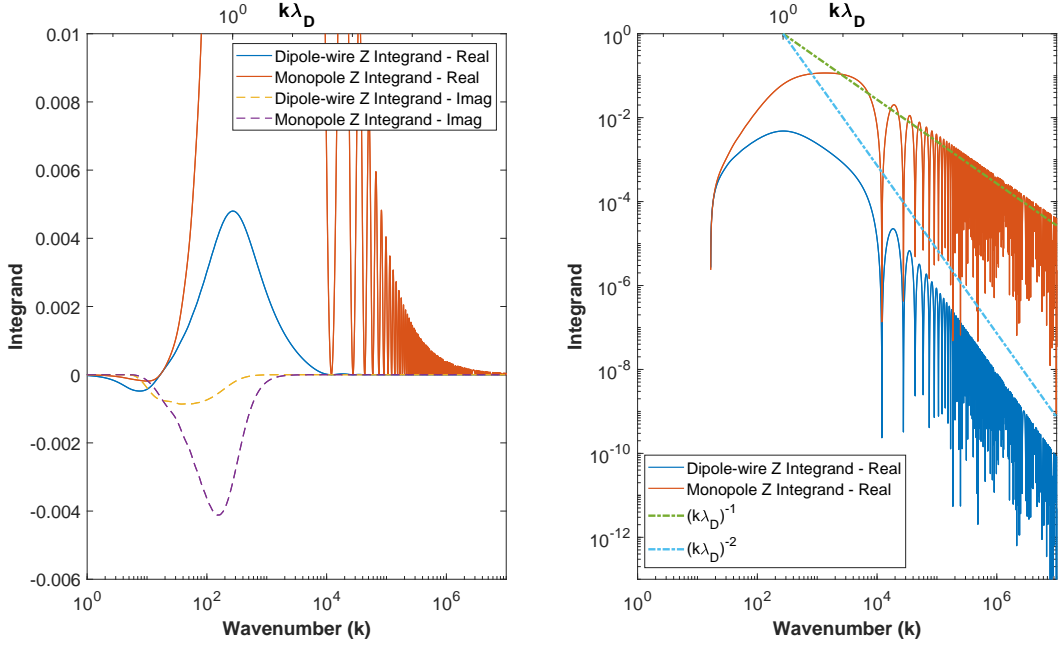}
\caption{Real and imaginary parts of the integrand, ${F_n(kL) J_0^2(ka)/\epsilon_L}$, plotted against $k$ (bottom axis) and $k\lambda_D$ (top axis) for (\ref{eq:Imped_w/F1}) and (\ref{eq:Imped_Mon1_w/F3}), corresponding to a wire dipole antenna and a monopole antenna for $n$ $=$ 1 or $m1$, respectively. The log-linear panel (left) reveals the form of the mostly negative imaginary parts while the log-log panel (right) includes additional $(k\lambda_D)^{-1}$ and $(k\lambda_D)^{-2}$ lines to show the rate of decrease for the real parts of the integrand at large $k$. In both panels $L$ $=$ \SI{0.3}{\meter}, $a$ $=$ \SI{2e-4}{\meter}, $\omega$ $=$ \SI{3.5e6}{\radian\per\second}, and average ionospheric conditions at \SI{300}{\kilo\meter} altitude are used, i.e. $n_e$ $=$ \SI{5.84e11}{\per\cubic\meter} and $T_e$ $=$ \SI{1690}{\kelvin}. Only the positive values of the imaginary part of the integrand in the log-log panel (right) are plotted.}
\label{fig:Integ_vs_k}    
\end{figure}

Figure \ref{fig:Integ_vs_k} compares how the integrands in the antenna impedances $Z_a$ given by (\ref{eq:Imped_w/F1}) and (\ref{eq:Imped_Mon1_w/F3}) vary with the antenna response function used. To calculate the integrand, average ionospheric conditions at \SI{300}{\kilo\meter} altitude were used, with $n_e = \SI{5.84e11}{\per\cubic\meter}$ and $T_e = \SI{1690}{\kelvin}$. A value of $\omega = \SI{3.5e6}{\radian\per\second}$ was chosen as a sample value for $\omega$, the antenna length $L = \SI{0.3}{\meter}$ and antenna radius $a = \SI{2e-4}{\meter}$. Note that $L/\lambda_D = 80.8$ and $\omega/\omega_p = 0.08$. 

Figure \ref{fig:Integ_vs_k} shows that the real parts of both integrands have a peak near $k\lambda_D \approx 1-3$. While the dipole peak at $k\lambda_D = 1$ is quite sharp and prominent, the peak is much broader and larger in the monopole case - almost 25 times larger, as seen in the log-log plot. The real parts of both integrands converge to zero for large $k$, but this happens much faster for the dipole case. The monopole function oscillates towards zero, decreasing approximately as $k^{-1}$, which is clear in the log-log plot with the $(k\lambda_D)^{-1}$ relationship overlaid. The dipole case also oscillates towards zero at large $k$ (due to the Bessel function), decreasing approximately as $k^{-2}$. 

The imaginary parts of the integrands in Figure \ref{fig:Integ_vs_k} are also similar in their behavior but with some differences. Both become negative near $k \approx$ \SI{10}{\per\meter} ($k\lambda_D \approx 0.1$) and have a negative peak near $k \simeq $ \SIrange{50}{200}{\per\meter} ($k\lambda_D \simeq  0.2 - 0.8$). The dipole case has a flatter peak while the monopole peak is much more prominent, in fact being very similar in shape to the real part of the dipole function. The minimum value in both cases is reached below $k\lambda_D = 1$.

\begin{figure}
\includegraphics[width=\textwidth]{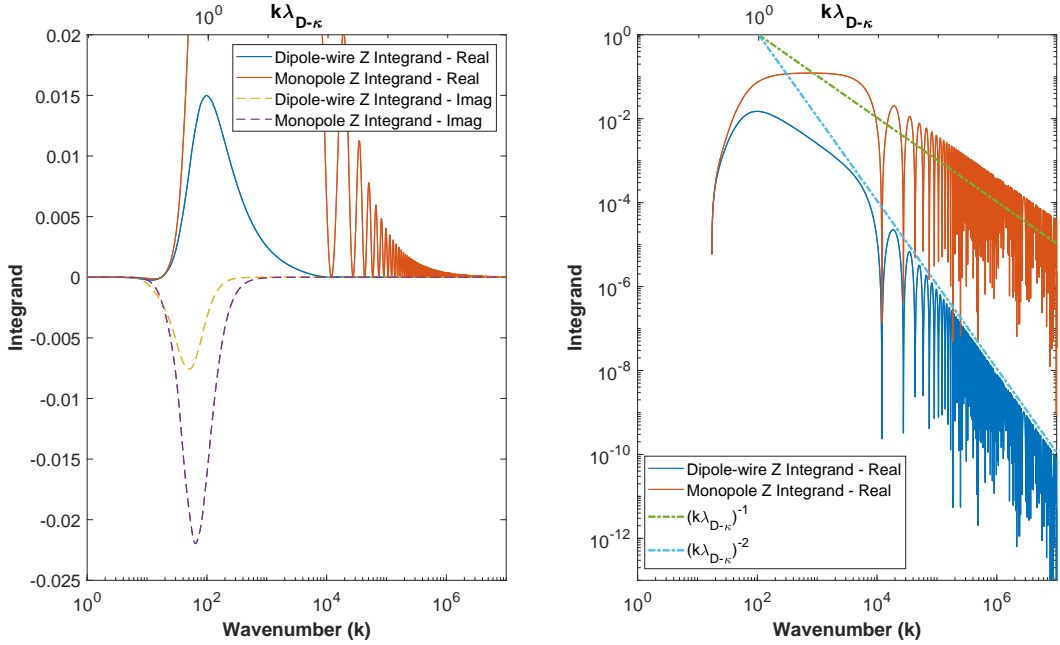}
\caption{Real and imaginary parts of the integrand, ${F_n(kL) J_0^2(ka)/\epsilon_L}$, plotted against $k$ (bottom axis) and $k\lambda_D$ (top axis), assuming a kappa distribution. This means that (\ref{eq:epsL_kap}) was used for $\epsilon_L$, rather than (\ref{eq:epsL}) as in Figure \ref{fig:Integ_vs_k}, and (\ref{eq:LD_kap}) is used to define the Debye length, with the notation also changing to $\lambda_{D-\kappa}$. Otherwise the same variables, conditions and format are used as in Figure \ref{fig:Integ_vs_k}. }
\label{fig:Integ_vs_k_kap}    
\end{figure}

Figure \ref{fig:Integ_vs_k_kap} shows the effect of changing the VDF from a Maxwellian to a kappa distribution. With $\kappa = 4$ the same integrand was calculated and the real and imaginary parts plotted in log-linear and log-log panels. The Debye length $\lambda_{D-\kappa}$ is now defined by (\ref{eq:LD_kap}). The properties of the functions are very much the same in Figures \ref{fig:Integ_vs_k} and \ref{fig:Integ_vs_k_kap}, showing that changing to a kappa distribution intorduces no qualitative differences. In detail, the real parts show a smooth rise to positive values and then an oscillatory decrease towards zero, while the imaginary parts show smooth non-oscillatory behavior with a peak near $k\lambda_{D-\kappa} \approx 1-3$. One difference for the Maxwellian case is that the peaks are shifted to slightly higher $k$, which is important only around the plasma peak where the pole of integration is located. A prominent and sharp peak is reached for the real part of the dipole integrand and imaginary part of the monopole integrand, albeit reaching a smaller positive and larger negative height than in Figure \ref{fig:Integ_vs_k}, respectively.

\begin{figure}
\includegraphics[width=\textwidth]{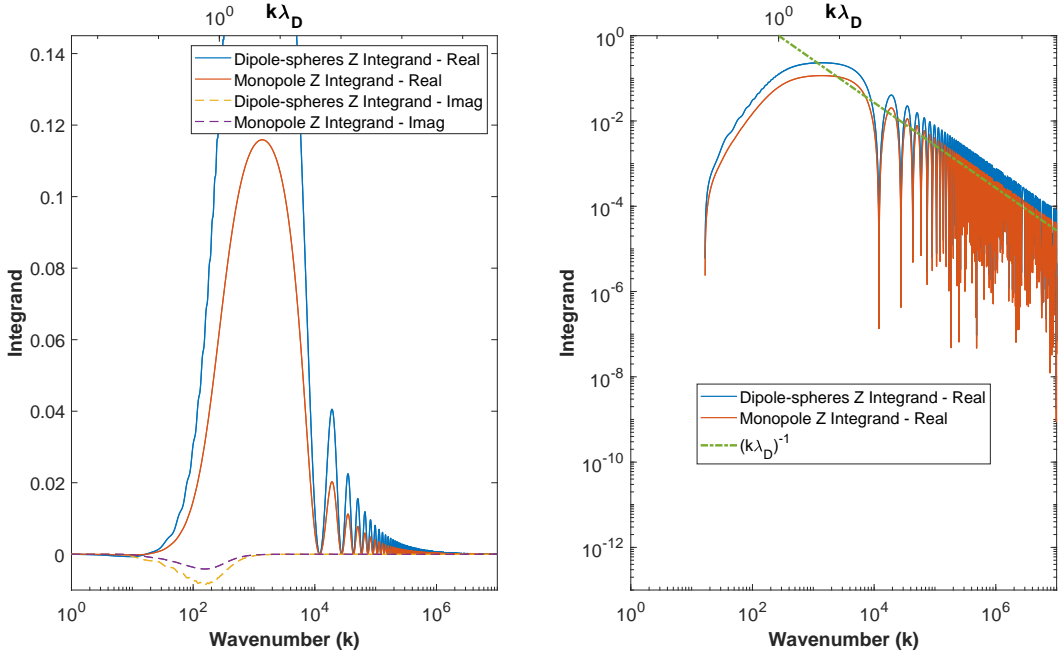}
\caption{Same as Figure \ref{fig:Integ_vs_k} but comparing the monopole response with a double-sphere dipole antenna rather than a wire dipole. }
\label{fig:Integ_vs_k_sph}    
\end{figure}

Figure \ref{fig:Integ_vs_k_sph} uses the double-sphere antenna response function in place of the function for the wire dipole case in Figure \ref{fig:Integ_vs_k}, so that we can clearly see the very strong similarities between the integrals in (\ref{eq:F3}) and (\ref{eq:F_sph}) for the monopole and double-sphere dipole cases. The factor of 2 difference in the leading constant for the expressions is clearly seen in Figure \ref{fig:Integ_vs_k_sph}, with the double-sphere values roughly double their monopole counterparts. Most importantly, the general shapes of the functions are effectively identical - for the real part a large broad peak followed by the oscillatory decrease to zero as $\approx k^{-1}$ while the imaginary parts show a single smooth negative peak or trough. The small wiggles for $k$ values below the peak in the double-sphere dipole integrand is one feature that differs between the two cases.

\begin{figure}
\centering
\includegraphics[width=\linewidth]{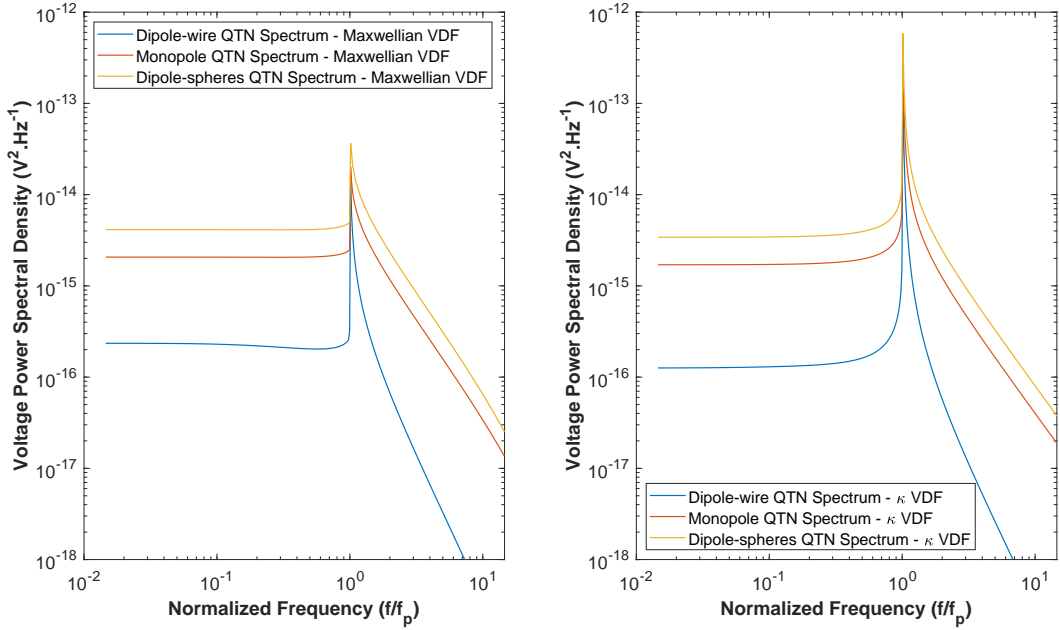}
\caption{Comparison of the quasi-thermal noise (QTN) voltage power spectrum predicted for dipole and monopole antennas using (\ref{eq:QTN_gen}) and (\ref{eq:Imped_w/F}) assuming a (left) Maxwellian VDF and (right) kappa VDF. The wire dipole case uses (\ref{eq:Imped_w/F1}), the monopole uses (\ref{eq:Imped_Mon1_w/F3}) and the double-sphere dipole uses (\ref{eq:F_F_sph_sin}) in (\ref{eq:Imped_w/F}). Parameters used are $L$ $=$ \SI{0.3}{\meter}, $a$ $=$ \SI{2e-4}{\meter}, $n_e$ $=$  \SI{5.84e11}{\per\cubic\meter}, $T_e$ $=$ \SI{1690}{\kelvin} and $\kappa$ $=$ $4$. Based on these conditions $f_p = \SI{6.9}{\mega\hertz}$.}
\label{fig:QTN_compare}
\end{figure}

\begin{figure}
\centering
\includegraphics[width=\linewidth]{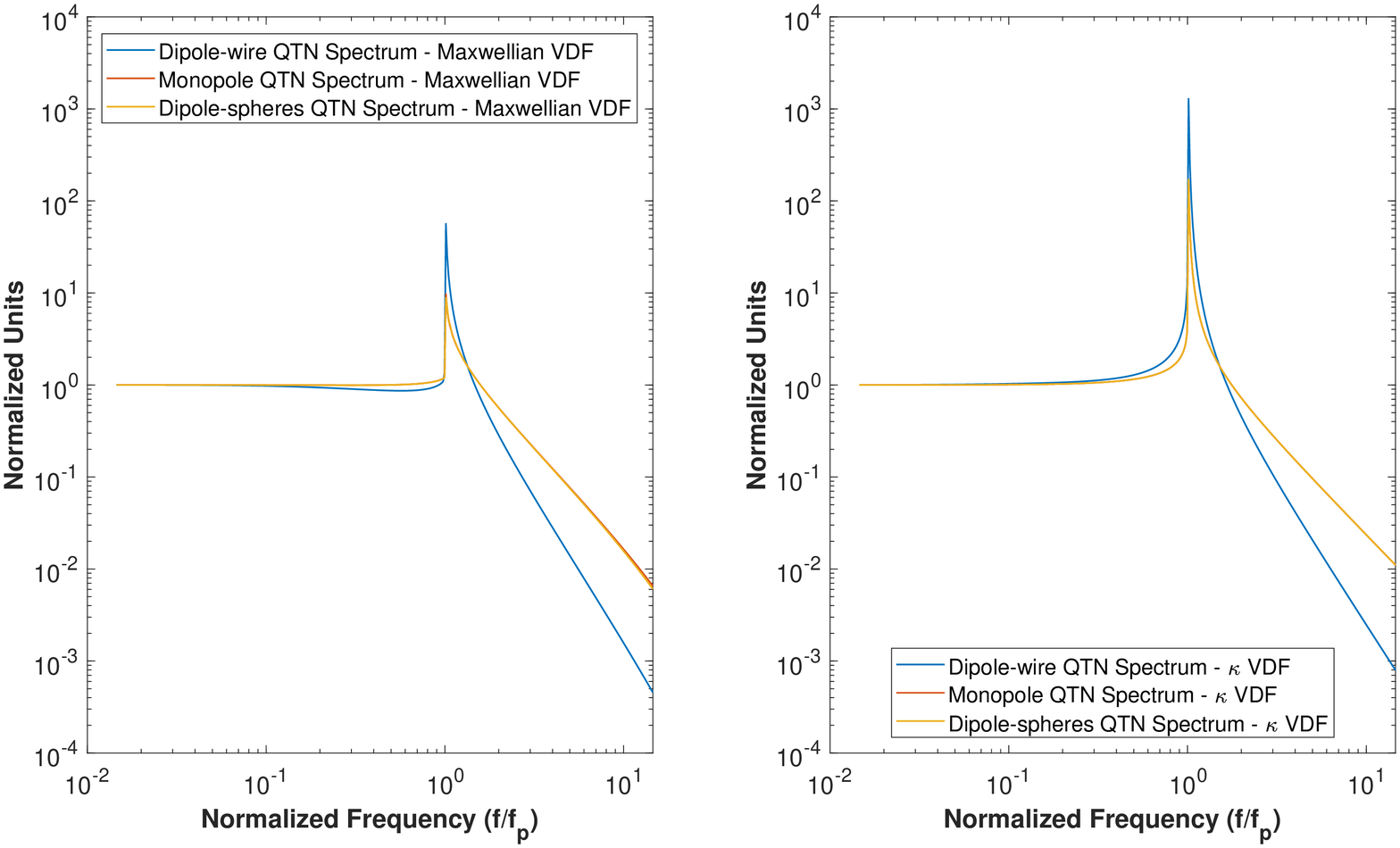}
\caption{Same spectra and conditions as in Figure \ref{fig:QTN_compare} but normalized to the QTN amplitude at $f = \SI{1e5}{\hertz}$ for $f_p = \SI{6.9}{\mega\hertz}$.}
\label{fig:QTN_compare_norm}
\end{figure}

Figure \ref{fig:QTN_compare} shows the predicted QTN spectra for the dipole and monopole antennas immersed in an ionospheric plasma with either a (left) Maxwellian or (right) kappa VDF with $\kappa = 4$. The antenna response functions in (\ref{eq:F1}), (\ref{eq:F3}) and (\ref{eq:F_sph}) are used for the wire dipole, monopole and double-sphere impedance $Z_a$, respectively, in (\ref{eq:QTN_gen}) and (\ref{eq:Imped_w/F}). The plasma conditions used are averages for the ionosphere at \SI{300}{\kilo\meter} above Earth's surface obtained from the IRI model as used in Figures \ref{fig:Integ_vs_k} - \ref{fig:Integ_vs_k_kap}. 

The wire dipole spectrum (blue) is relatively flat at about \SI{2.35e-16}{\volt\squared\per\hertz} for frequencies below the plasma frequency $f_p =$ \SI{6.86e6}{\hertz} and then rises to a peak just above $f_p$. Comparing this with the approximation in Table 1 of \citet{meyer1989}, for $f \ll f_p$ the voltage power should be $V_0^2 = \num{5e-16}\times \sqrt{T_e}\times (\lambda_D / L) \hspace{5pt} \SI{}{\volt\squared\per\hertz}$. For the conditions used in this paper, $V_0^2 = \SI{2.54e-16}{\volt\squared\per\hertz}$, which is within 8\% of the calculated value and therefore our calculation agrees well with past literature. For frequencies greater than $f_p$ the spectrum has a $f^{-3.2}$ relationship, which was calculated using the curve fitting tool in Matlab assuming a power law ($ax^b$) and fitted for values from $f/f_p = 2.6$ onward (with 95\% confidence the exponent is between -3.21 and -3.22). The $\approx f^{-3}$ relationship above $f_p$ is well known in the literature for wire dipole antennas \citep{meyer1989}. For the kappa distribution (gold), the wire dipole spectrum also starts from a similar level for $f \ll f_p$ but the peak is not as prominent. The fall off for large $f$ follows a very similar relationship as for the Maxwellian, $f^{-3.2}$, with 95\% confidence the exponent is between -3.14 and -3.16.

The monopole spectrum (red) is also relatively flat for low frequencies, but with a magnitude of \SI{2.07e-15}{\volt\squared\per\hertz} that is a factor of 10 higher than for the wire dipole case, and rises to a slightly higher peak just above $f_p$. However, the calculated fall-off for $f \gg f_p$ is different and the spectrum is proportional to $f^{-2.2}$ (with 95\% confidence the exponent is between -2.18 and -2.17). Interestingly, a fall-off of $\approx f^{-2}$ is well known for double-sphere antennas \citep{meyer1989}. In the kappa case, the monopole fall-off is $f^{-2.1}$, lying between -2.08 and -2.09 with 95\% confidence. The double-sphere dipole spectrum (gold) is very similar to the monopole, although a factor of 2 higher, as seen in both the left and right panels of Figure \ref{fig:QTN_compare}. The value of the flat spectrum for $f < f_p$ (\SI{4.13e-15}{\volt\squared\per\hertz}) also agrees within 1\% of the approximation from \citet{meyer1989} of $V_0^2 = m_ev_T / (\pi^{3/2}\epsilon_0) = \SI{4.18e-15}{\volt\squared\per\hertz}$ for $f\ll f_p$. The relationship for large values of $f$ is effectively the same as the monopole case with the same confidence levels.

The similarities between the monopole and double-sphere spectra are even more obvious in Figure \ref{fig:QTN_compare_norm}, which shows the spectra normalized to their value at $f = \SI{1e5}{\hertz}$. Both the double-sphere and monopole spectra overlap almost exactly for the Maxwellian case at all frequencies. For the kappa distribution there is a small ($\simeq 10 - 20\%$) but noticeable difference between the height of the two sets of spectra at the peak above $f_p$ - the double-sphere spectrum is slightly larger, which is expected due to the suprathermal electrons of a kappa distribution \citep{chateau1991, chat2009}. However, the shape and variations of the QTN spectra for the monopole and double-sphere antennas are extremely similar.

Figure \ref{fig:QTN_compare_norm} also shows the large prominence of QTN peak above $f_p$ of the wire dipole spectrum compared to the double-sphere dipole and monopole cases. For the Maxwellian VDF, the QTN peak rises to $\approx 70$ times the flat spectrum value (at $f \ll f_p$) for the wire dipole whereas the ratio is only $\simeq 10$ for the monopole and double-sphere dipole spectra. In the kappa distribution case, the wire dipole peak only rises to about 20 times the flat spectrum level compared to a ratio $\simeq 4 -5$ for the monopole and double-sphere spectra. The faster fall off rate at $f \gg f_p$ for the wire dipole spectrum is also clear in Figure \ref{fig:QTN_compare_norm}. 

The height and width of the QTN peaks would affect the expected voltage signal at the receiver as the measured spectra involve some integration over frequency - the higher and wider the peak the larger the expected voltage signals. The peak prominence is generally determined by $L/\lambda_D$, as described in \citet{couturier1981}; for $L/\lambda_D \leq 1$ the spectrum is almost completely flat but as the $L/\lambda_D$ becomes larger, the peak just above $f_p$ becomes higher and sharper. In Figures \ref{fig:QTN_compare} and \ref{fig:QTN_compare_norm} $L/\lambda_D = 80.8 \gg 1$, explaining the prominence of the peaks seen (especially for the wire dipole).


\begin{figure}
\centering
\includegraphics[width=\linewidth]{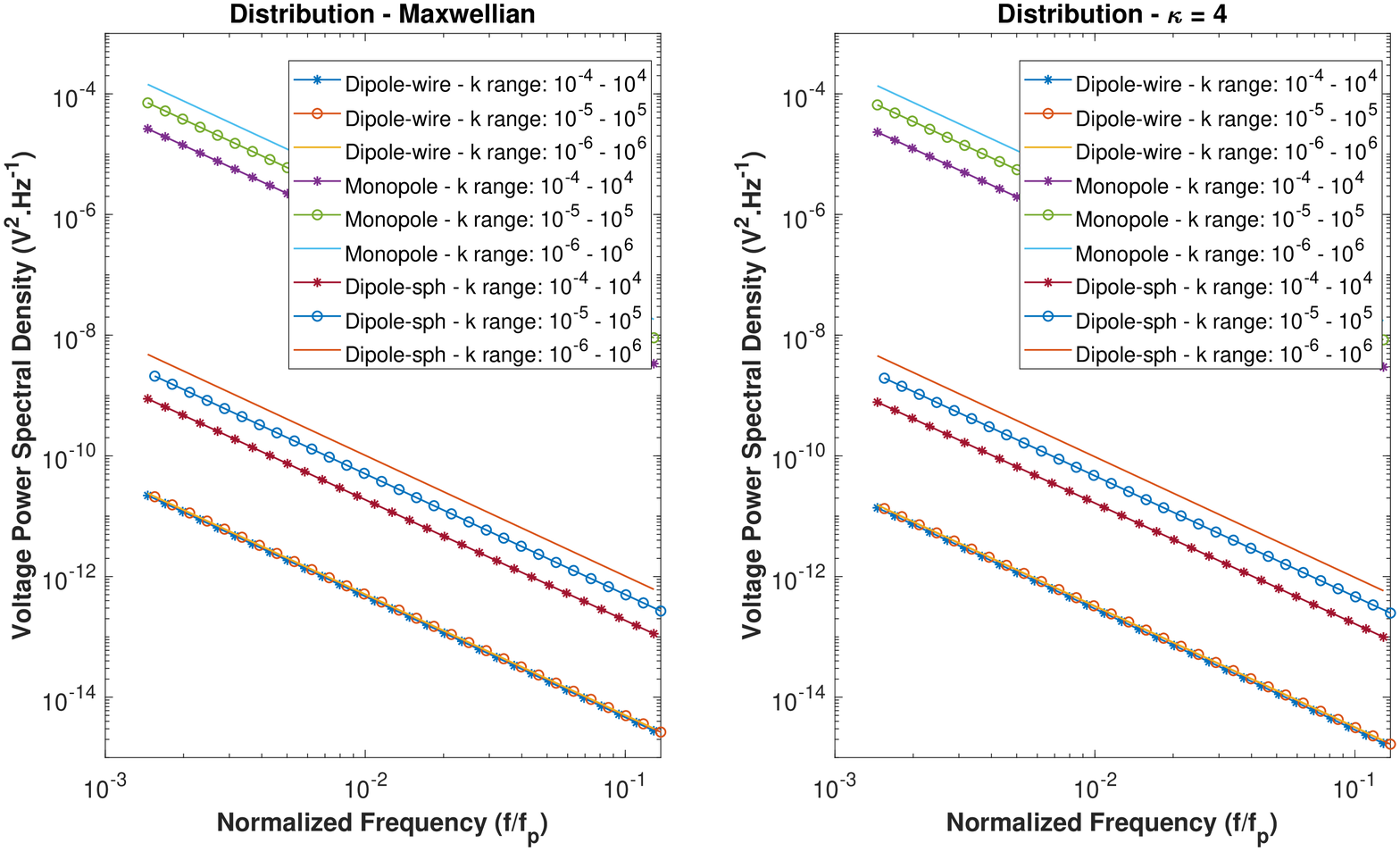}
\caption{Comparison of the shot noise voltage power spectra for wire dipole, double-sphere (dipole-sph) and monopole antenna configurations predicted using either (\ref{eq:F1}), (\ref{eq:F_sph}) or (\ref{eq:F3}), respectively, and (\ref{eq:Shot_gen}) and (\ref{eq:Imped_w/F}) for various ranges of $k$ and (left) Maxwellian and (right) kappa ($\kappa$$=$$4$) VDFs for $L$ $=$ \SI{0.3}{\meter}, $a$ $=$ \SI{2e-4}{\meter}, $n_e$ $=$ \SI{5.84e11}{\per\cubic\meter}, and $T_e$ $=$ \SI{1690}{\kelvin}. The three spectra for each antenna configuration show the various $k$ ranges identified in the inset over which (\ref{eq:Imped_w/F}) is integrated. Values of $k$ are in \SI{}{\per\meter}.}
\label{fig:Shot_compare}
\end{figure}

Figure \ref{fig:Shot_compare} shows a series of shot noise spectra for dipole and monopole antenna configurations, predicted using (\ref{eq:Shot_gen}) with different $k$ integration limits for both Maxwellian and kappa VDFs. Again (\ref{eq:F1}), (\ref{eq:F3}) and (\ref{eq:F_sph}) are used for the antenna response functions in (\ref{eq:Shot_gen}) but the surface area $S$ is introduced as an additional parameter through 
\begin{linenomath*}
\begin{equation}
\label{eq:N_e_def}
N_e = 1/{\sqrt{4\pi}}n_e v_T S. 
\end{equation}
\end{linenomath*}
For the dipole antennas this is the surface area of the two antenna arms while for the monopole antenna this is the surface area of the spacecraft body. This is because a monopole measures the potential difference between the antenna arm and the spacecraft body and the spacecraft area is assumed to be much larger than the antenna. In this paper we assume that a 1U CubeSat (\SI{10}{\centi\meter} $\times$ \SI{10}{\centi\meter} $\times$ \SI{10}{\centi\meter} sized spacecraft) is connected to the monopole antenna. For the antenna parameters we are considering, $a = \SI{2e-4}{\meter}$ and $L=\SI{0.3}{\meter}$, this means that $S$ is about 160 times larger for the monopole than for the dipole case. However, the fact that a monopole rather than a dipole antenna is used affects the theoretical shot noise level as well - the total noise for a dipole should be twice that for one arm \citep{meyer1983}. Therefore, from this difference in $S$ we expect for the monopole shot noise spectra to be larger by about a factor of 80, all else being equal, than the dipole spectra. In fact, many space missions that carry on-board both a dipole and monopole antenna (or option for switching between the two configurations) have seen increased noise levels in the monopole case. This effect has been seen in comparisons of dust impacts of the PRA (monopole) and PWS (dipole) instruments on board the Voyager 2 spacecraft \citep{mann2011} and for the Wind and STEREO spacecraft antennas \citep{meyer2009b}. However, there is still debate in the literature on the exact explanation for these variations \citep{gurnett1983, tsintikidis1994, meyer2014, ye2016, maj2018}. 

Figure \ref{fig:Shot_compare} uses the same ionospheric conditions as Figures \ref{fig:QTN_compare} and \ref{fig:QTN_compare_norm}. For each antenna configuration spectra are shown for three different ranges of $k$ in the integrals in (\ref{eq:Imped_w/F}). It is clear at a glance of Figure \ref{fig:Shot_compare} that the predicted shot noise spectra are convergent for the wire dipole case but not for the monopole or double-sphere dipole cases. In detail, for the wire dipole case increasing the integration range from $k = 10^{-4}$ - $10^{4}$ \SI{}{\per\meter} to $10^{-6}$ - $10^{6}$ \SI{}{\per\meter} brings the shot noise level only very slightly higher but with clear convergence to the final spectrum. For the monopole case, however, increasing the $k$ integration range increases the shot noise level considerably and there is no sign of convergence occurring; from $k = 10^{-4}$ - $10^{4}$ \SI{}{\per\meter} to $10^{-6}$ - $10^{6}$ \SI{}{\per\meter} the level increases by almost an order of magnitude. This is expected from Figure \ref{fig:Integ_vs_k} because the real part of the integrand does not converge quickly enough towards zero as it has a $k^{-1}$ functional form, this leading to only logarithmic convergence. This behavior also occurs for the double-sphere antenna, as seens in Figure \ref{fig:Integ_vs_k_sph} and \ref{fig:Shot_compare}. 

We can also see in Figure \ref{fig:Shot_compare} that the effects of changing the VDF to a kappa distribution from a Maxwellian are minimal. In detail, for these parameters the shot noise spectrum level decreases by less than a factor of 2 for the wire-dipole case while the monopole and double-sphere spectra vary even less. The similarities in the double-sphere and monopole antenna response functions in (\ref{eq:F3}) and (\ref{eq:F_sph}) and Figures \ref{fig:Integ_vs_k_sph} - \ref{fig:QTN_compare_norm} thus predictably carry over to produce very similar shot noise spectra in Figure \ref{fig:Shot_compare} when the integration limits are changed, albeit separated by five orders of magnitude. Therefore increasing $k$ merely increases the final integral and does not allow the use of the monopole antenna response function (\ref{eq:F3}) or the well established double-sphere expression (\ref{eq:F_sph}) to predict the shot noise level. 

This may be caused by the implementation used in this paper and a more reasonable result could be achieved by considering the $\Delta V_\omega^2$ term (known as the correction term) described in \citet{meyer1983}. In \citet{meyer1983} the author thoroughly derives the voltage power spectrum for finite width dipole antennas in a plasma which is separated into QTN and shot noise terms plus an additional $\Delta V_\omega^2$ term. This additional term is ignorable if $\omega \times \min(L,\lambda_D)/v_T <1$ and $\omega \neq \omega_p$ and the frequency range used here may be too close to $f_p$. However, calculations with $f = $\SIrange{1}{1000}{\hertz} do not lead to convergence either. It should be noted that the literature seems to be lacking in predictions of the shot noise power spectra for double-sphere antennas and this may be a wider issue.


 \begin{table}
 \caption{Effect of changing the integration limits on the calculated capacitance (\ref{eq:Capacit_imped}) which uses the impedance to calculate capacitance}
 \label{tab:capac_integ_lim}
 \centering
 \begin{tabular}{c c c c}
 \hline
  $k$ Integration Limits  & Monopole Capacitance  & Double-sphere Capacitance & Wire Dipole  \\
 \hline
   $10^{-4}$ - $10^{4}$ \SI{}{\per\meter}  & \SI{3.33e-14}{\farad} &
   \SI{1.66e-14}{\farad} &
   \SI{2.88e-14}{\farad}\\
   $10^{-5}$ - $10^{5}$ \SI{}{\per\meter}  & \SI{2.03e-14}{\farad} & 
   \SI{1.01e-14}{\farad} &
   \SI{2.77e-14}{\farad}\\
   $10^{-6}$ - $10^{6}$ \SI{}{\per\meter} &
   \SI{1.43e-14}{\farad} &
   \SI{7.13e-15}{\farad} &
   \SI{2.76e-14}{\farad} \\
   $10^{-7}$ - $10^{7}$ \SI{}{\per\meter} &
   \SI{1.10e-14}{\farad} &
   \SI{5.49e-15}{\farad} &
    \SI{2.76e-14}{\farad}
\end{tabular}
\end{table}

The capacitances for the monopole, double-sphere and wire dipole antennas were calculated using (\ref{eq:Capacit_imped}) and the impedances calculated for the QTN and shot noise with a Maxwellian VDF - the results are presented in Table \ref{tab:capac_integ_lim}. For the wire dipole case the integration limits do not change the impedance value appreciably and therefore we can quote the capacitance result \SI{2.76e-12}{\farad} for the $k$ range $10^{-6}$ - $10^{6}$ \SI{}{\per\meter} as a `final' result (computationally faster than the $10^{-7}$ - $10^{7}$ \SI{}{\per\meter} integration limit considered). However, for both the monopole and double-sphere cases changing the range of $k$ integration changes the capacitances significantly, by a factor of $\approx 3$ for our cases. Furthermore, the low frequency approximation for the capacitance of a wire dipole antenna from (\ref{eq:Capacit_dipole}) yields \SI{2.86e-12}{\farad}, which is within 4\% of the value calculated from (\ref{eq:Capacit_imped}). The same approximation for the double-sphere antenna yields \SI{1.11e-14}{\farad}, which is within 10\% of the value calculated above with $k$ limits of $10^{-5}$ - $10^{5}$ \SI{}{\per\meter} but further increases in the $k$ range lead to a clear reduction in the calculated capacitance. Finally, the approximation $C_{monopole} = 2C_{dipole} = \SI{5.71e-12}{\farad}$ for the monopole capacitance is over 2 orders of magnitude larger than the values calculated from (\ref{eq:Capacit_imped}) and the discrepancy becomes larger the greater the range of $k$ that is used. Thus the values predicted for the capacitances yield several problems associated with the use of the antenna response function (\ref{eq:F3}).

\section{Discussion}
\label{sec:Discuss}
The predicted QTN spectrum, as presented in Figure \ref{fig:QTN_compare}, is one case where our newly derived monopole response function provides a reasonable, converged prediction which can be tested and used for measurement in the real-world. The reason is that unlike the predictions for the shot noise and capacitance made using this new function, increasing the range of $k$ in the impedance integral for the monopole case does not impact on the final result. This is due to the fact that in for the QTN only the real part of $Z_a$ matters if we disregard gain effects and the factor of $i$ outside the integral in the impedance (\ref{eq:Imped_Mon1_w/F3}) implies that it is the imaginary component of the integrand that contributes to the real part of $Z_a$. In Figure \ref{fig:Integ_vs_k} the imaginary component for the monopole case is shown to be well behaved and although it has a larger magnitude and sharper peak than the wire dipole case it follows the same functional form and crucially approaches zero quickly enough at large $k$ ($\propto k^{-2}$), being visually close to zero by $k=$\SI{1000}{\per\meter}. This ensures that the integral is finite when integrating up to higher $k$ values. Therefore (\ref{eq:F3}) may be useful for predicting the monopole QTN spectrum, both for Maxwellian and kappa VDFs; however, comparison with observational data is required to verify this.

The other applications of the wire dipole, double-sphere and monopole antenna response functions in Section \ref{sec:Compare}, however, reveal a number of issues with the newly derived monopole response function. Although the same procedure was used to derive both the wire dipole and monopole functions, there are a number of features that are problematic for the monopole case. Firstly, Figure \ref{fig:Fn_vs_x} reveals the non-zero and almost constant nature of the monopole function for large $x = kL$, which in Figure \ref{fig:Integ_vs_k} is seen to affect the real part of the integrand used in the impedance integral (\ref{eq:Imped_Mon1_w/F3}). At large $k \gtrsim$ \SI{e4}{\per\meter}, the real component of the monopole integrand oscillation rises to a much larger peak and does not fall off as quickly as the dipole integrand, in fact falling off $\propto k^{-1}$ instead of $\propto k^{-2}$. This is problematic when integrating from $k=0$ to $k=\infty$ (or choosing numerical limits close to these values) as the final result only converges logarithmically. This is evident in Figure \ref{fig:Shot_compare} where we have used increasingly larger $k$ ranges for the impedance integral in order to determine the level of shot noise. Importantly, all these features are in common with the double-sphere antenna, which has a widely cited antenna response function very similar in functional form to the monopole response function derived in this paper. The reason for these problems is the $k^{-1}$ fall-off of the integrand's real part, so that the integral does not converge as $k\rightarrow \infty$.



\begin{figure}
\centering
\includegraphics[width=\linewidth]{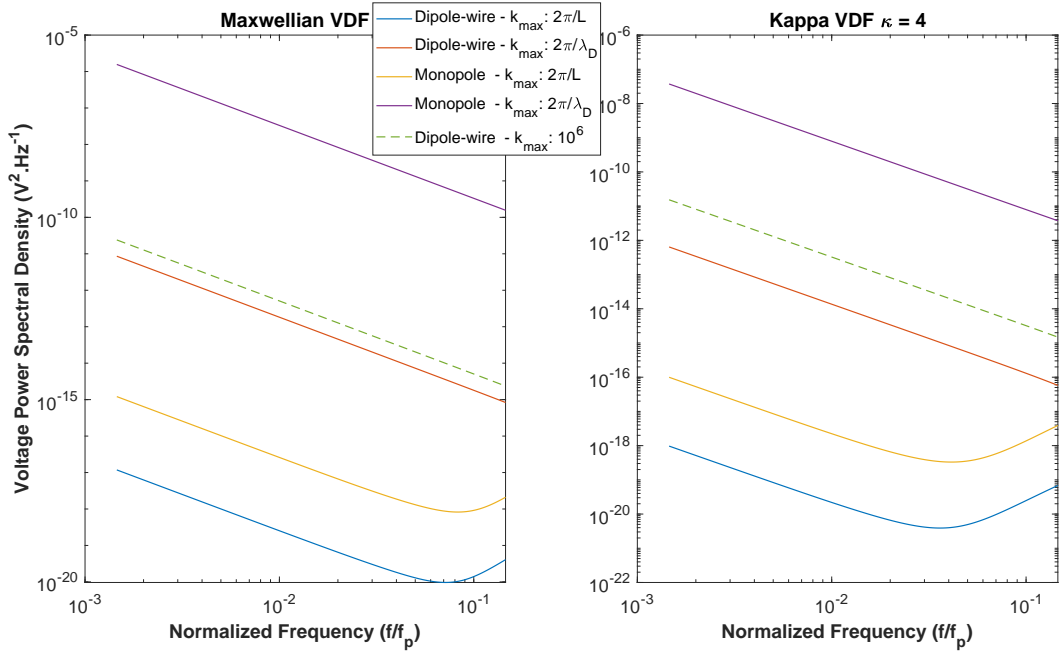}
\caption{Comparison of shot noise spectra for a wire dipole and monopole antenna using the same equations and conditions as Figure \ref{fig:Shot_compare} but varying the maximum $k$ reached in the integrals (\ref{eq:Imped_w/F1}) and (\ref{eq:Imped_Mon1_w/F3}) between $k_{max}$ $=$ $2\pi/L$ $=$ \SI{21}{\per\meter}, $k_{max}$ $=$ $2\pi/\lambda_D$ $=$ \SI{1700}{\per\meter}, and $k_{max}$ $=$ \SI{e6}{\per\meter}. The minimum $k$ used in the integral in all cases is the same: $k_{min}$ $=$ \SI{e-6}{\per\meter}. Two VDFs are presented: (left) Maxwellian and (right) kappa. For the kappa case, $\lambda_{D-\kappa}$ rather than $\lambda_D$ is used for the Debye length.}
\label{fig:kmax_compare}
\end{figure}

Restricting $k$ on physical grounds is a possible solution to this infinite integral problem. One way is to use the Debye length as a limiting factor, i.e. setting $k_{max} = 2\pi/\lambda_D$, which would imply that the wavelength cannot be smaller than the Debye length. Another way is to use the antenna length itself and set $k_{min} = 2\pi/L$, implying that the length of the antenna constrains the wavelength that is detected to the case of a single wavelength fitting across the length of one antenna arm. Figure \ref{fig:kmax_compare} compares the wire dipole and monopole predictions for each of $k_{max} =$ $2\pi/L$ and $2\pi/\lambda_D$, as well as comparing these against the dipole shot noise spectrum for the $k$ range \num{e-6} - \num{e6} \SI{}{\per\meter} from Figure \ref{fig:Shot_compare}, represented in Figure \ref{fig:kmax_compare} as the dashed curve. We have also performed this comparison for both Maxwellian and kappa VDFs, showing that the results are similar and that the VDF does not affect the final result. For the dipole antenna case, it is clear that restricting $k_{max}$ does not recreate the same result as seen in Figure \ref{fig:Shot_compare}. With $k_{max} = 2\pi/\lambda_D = \SI{1.69e3}{\per\meter}$ the spectrum is 2.8 times lower than the dashed curve while with $k_{max} = 2\pi/L = \SI{20.9}{\per\meter}$ the spectrum is not only \num{2e6} times lower but the simple power law relationship for $V^2(f)$ breaks down at higher frequencies. This shows that the majority of the response comes from the range $2\pi/L - 2\pi/\lambda_D$ for the wire dipole case. For the monopole case with $k_{max} = 2\pi/L$, the spectrum has the same functional form but is 2 orders of magnitude closer to the dashed line (the wire dipole spectrum). For $k_{max} = 2\pi/\lambda_D$ the monopole spectrum is over 4 orders of magnitude larger than the dashed dipole spectrum, and as seen in Figure \ref{fig:Shot_compare} this behavior continues as $k_{max}$ is increased further. Again, a major contribution comes from the range $2\pi/L - 2\pi/\lambda_D$.

\begin{figure}
    \centering
    \includegraphics[width=\linewidth]{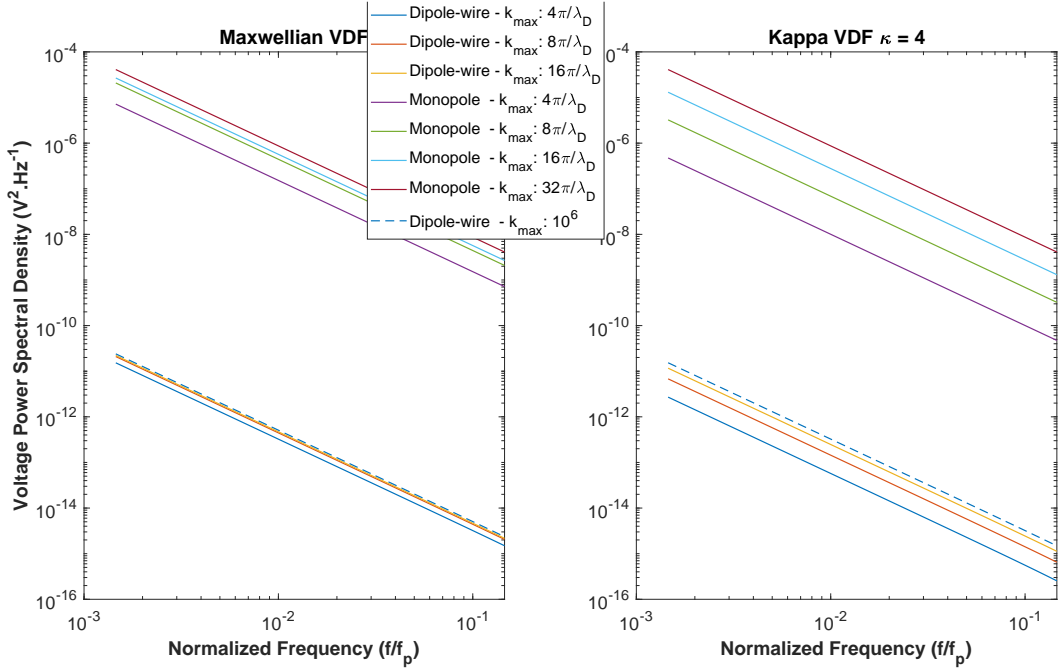}
    \caption{Comparison of shot noise spectra as in Figure \ref{fig:kmax_compare} but using $k_{max}$ $=$ $n$ $\times$ $(2\pi/\lambda_D)$ with $n$ $=$ $2, 4, 8$;
    $k_{max}$ $=$ $\num{e6}$ is also included. $k$ is in \SI{}{\per\meter} and $\lambda_{D-\kappa}$ rather than $\lambda_D$ is used in the kappa VDF case.}
    \label{fig:kmax_n2pi-LD}
\end{figure}

\begin{figure}
    \centering
    \includegraphics[width=\linewidth]{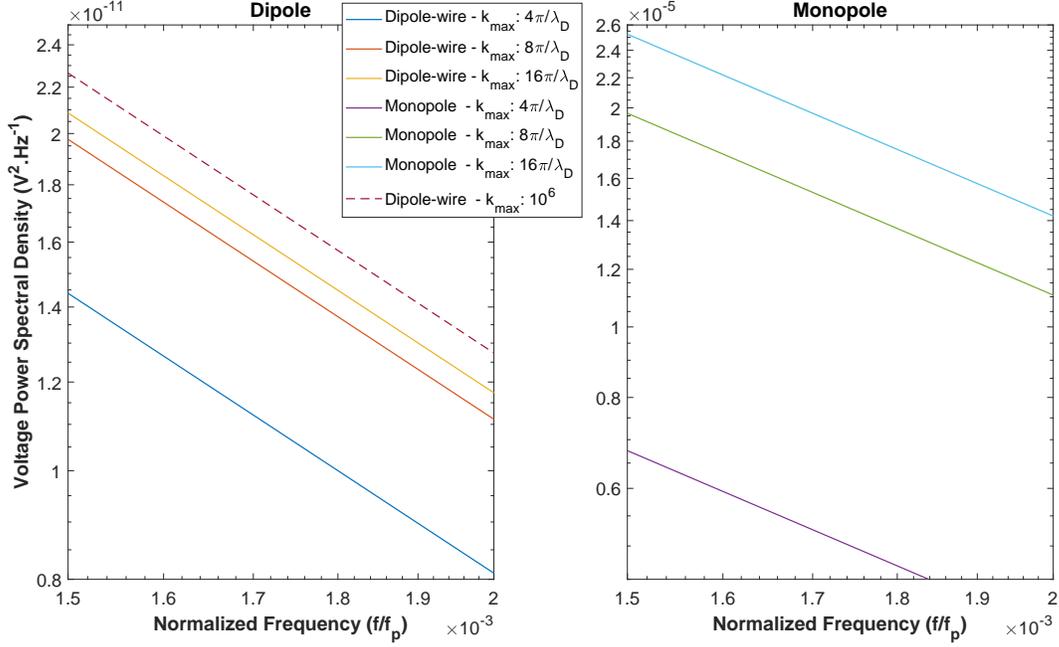}
    \caption{Zoomed-in version of the left, or Maxwellian, panel of Figure \ref{fig:kmax_n2pi-LD} showing the (left) wire dipole and (right) monopole predictions for shot noise.}
    \label{fig:kmax_n2pi-LD_zoom}
\end{figure}

The Debye length is the distance over which the electric potential decreases by a factor $1/e$ and is used to approximate when a charged object is shielded from the effect of charges further away. Therefore, plasma waves with wavelengths shorter than about the Debye length will be damped by the charges in the plasma. However, restricting the wavelength $\lambda$ to only one Debye length (or $k_{max} = 2\pi/\lambda_D$) may not be enough to integrate over all physically significant wavelengths - a factor $1/e$ is still $37\%$ of the initial charge. In Figures \ref{fig:kmax_n2pi-LD} and \ref{fig:kmax_n2pi-LD_zoom} we show the predictions for increasing $k_{max}$ to $2\times, 4\times$ and $8\times$ $2\pi/\lambda_D$, equivalent to decreasing the wavelength to $1/2, 1/4$ and $1/8$ of $\lambda_D$. 

In the dipole case, as expected increasing $k_{max}$ brings it closer to its final convergent shot noise level - at $k_{max} = 16\pi/\lambda_D$ the shot noise spectrum is $92\%$ ($76\%$) of the result for $k_{max} = \num{e6}$ for the Maxwellian VDF (kappa VDF). Increasing $k_{max}$ from $4\pi/\lambda_D$ to $8\pi/\lambda_D$ increases the shot noise level by $37\%$ ($150\%$) but from $8\pi/\lambda_D$ to $16\pi/\lambda_D$ the increase is only $5.6\%$ ($71\%$) for the Maxwellian VDF (kappa VDF). A further $8.5\%$ ($31\%$) increase on the $k_{max} = 16\pi/\lambda_D$ spectrum would bring it to the Maxwellian (kappa) spectrum for $k_{max} = \num{e6}$. 

For the monopole case, we initially see apparent convergence as in the dipole case but divergence at the highest $k_{max}$ calculated. Based on our calculations, increasing $k_{max}$ from $4\pi/\lambda_D$ to $8\pi/\lambda_D$ increases the shot noise level by $190\%$ ($580\%$) while from $8\pi/\lambda_D$ to $16\pi/\lambda_D$ the increase is $28\%$ ($300\%$) for the Maxwellian (kappa) VDF. A value of $k_{max} = 16\pi/\lambda_D$ implies that the wavelength is on a scale which is shielded to less than $5\%$ of the electric potential of undressed charges. However, increasing from $16\pi/\lambda_D$ to $32\pi/\lambda_D$ there is an increase in the spectrum level by $53\%$, greater than the $28\%$ increase from $8\pi/\lambda_D$ to $16\pi/\lambda_D$. Therefore, there may be a physically significant effect between these two values but ultimately the spectrum is not converging. A crucial point here is that with no obvious reference it is difficult to ascertain how close these values are to the "true" shot noise level. In the Maxwellian case, $16\pi/\lambda_D$ may be a sufficiently large value of  $k_{max}$ to account for most of the physically significant effects, especially based on the results for the dipole case (reaching $92\%$ of the final value) but further verification is necessary, experimental tests being best.

For the kappa VDF, it is clear in Figure \ref{fig:kmax_n2pi-LD} that convergence is slower than for the Maxwellian VDF. The dipole spectrum at $k_{max} = 16\pi/\lambda_D$ is only $76\%$ of the reference spectrum at $k_{max} = \num{e6}$ while the monopole spectrum increases by integer multiples of the previous levels when increasing $k_{max}$ from  $4\pi/\lambda_D$ to $8\pi/\lambda_D$ and then from $8\pi/\lambda_D$ to $16\pi/\lambda_D$. It may be difficult to achieve a specific kappa VDF in a laboratory setting to experimentally discover or verify the necessary $k_{max}$ limit, however this would again be the best approach. 

We may also choose the capacitance as a criteria for restricting $k_{max}$. Using the monopole impedance (\ref{eq:Imped_Mon1_w/F3}) calculated for $k_{max} = 2\pi/\lambda_D$ and $k_{max} = 16\pi/\lambda_D$ we find using (\ref{eq:Capacit_imped}) that $C_{mono2\pi/L} = \SI{1.36e-13}{\farad}$ and $C_{mono16\pi /\lambda_D} = \SI{2.67e-14}{\farad}$, respectively. These are over an order of magnitude away from the value calculated in Section \ref{sec:Compare}'s last paragraph, i.e. $C_{monopole} = 2C_{dipole} = \SI{5.71e-12}{\farad}$. If we use the low frequency approximation $C_{monopole}$ as the necessary criteria for determining the $k_{max}$ cut-off then we find that a value of $k_{max} = \SI{207}{\per\meter} = 0.122 \times (2\pi/\lambda_D)$ brings the calculated capacitance within 3\% of $C_{monopole}$. 

 \begin{table}
 \caption{Effect of restricting $k_{max}$ on capacitance results under plasma different conditions. Columns 4 and 5 are calculated using (\ref{eq:Capacit_imped}) and (\ref{eq:Capacit_dipole}), respectively. }
 \label{tab:restrict_k}
 \centering
 \begin{tabular}{c c c c c}
 \hline
  Altitude   & Density $n_e$  & Temperature $T_e$  & Capacitance with $k_{max}=0.12 \times \frac{2\pi}{\lambda_D}$  & Capacitance for $f\ll f_p$  \\
 \hline
   \SI{300}{\kilo\meter} & \SI{5.84e11}{\per\cubic\meter} & \SI{1.69e3}{\kelvin} & \SI{5.72e-12}{\farad} & \SI{5.71e-12}{\farad} \\
   \SI{800}{\kilo\meter}  & \SI{6.15e10}{\per\cubic\meter} & \SI{2.82e3}{\kelvin} & \SI{1.43e-12}{\farad} & \SI{3.88e-12}{\farad}   \\
   \SI{1500}{\kilo\meter}  & \SI{1.01e10}{\per\cubic\meter} & \SI{3.24e3}{\kelvin} & \SI{1.03e-12}{\farad} & \SI{3.17e-12}{\farad}  \\
\end{tabular}
\end{table}

To test this criterion for restricting $k_{max}$ under different plasma conditions we calculate the impedance with $k_{max} = 0.122 \times (2\pi/\lambda_D)$ under average ionospheric conditions at \SI{800}{\kilo\meter} and \SI{1500}{\kilo\meter} altitude from the IRI, as summarisized in Table \ref{tab:restrict_k}. The table also includes the IRI model values for $n_e$ and $T_e$ at \SI{800}{\kilo\meter} and \SI{1500}{\kilo\meter}, and a comparison against the low frequency approximation (\ref{eq:Capacit_dipole}).  
Although these values are closer than the orders of magnitude difference seen previously for the restrictions based on the length of the antenna and Debye length, these capacitances are still about a factor of 3 different from the predictions using the low frequency approximation. Also, using $k_{max}  = 0.122 \times (2\pi/\lambda_D)$ for the wire dipole case does not lead to reasonable capacitances either. For \SI{300}{\kilo\meter} altitude, a capacitance of \SI{3.71e-11}{\farad} is obtained, which is a factor of 13 larger than the dipole approximation, while at \SI{800}{\kilo\meter} and \SI{1500}{\kilo\meter} the capacitance is larger by a factor of 3.9 and 2.8, respectively, from the low frequency approximation.

Therefore the main problem with restricting $k_{max}$ is that a choice must be made on what this value should be. Basing it on physical grounds, such as restricting $k_{max}$ to $2\pi/\lambda_D$, does not recreate the same results as in Figure \ref{fig:Shot_compare} for the dipole antenna. Increasing this limit to 4 or 8 $\times 2\pi/\lambda_D$ will bring the spectrum closer but what this multiplicative factor should be is ultimately an arbitrary choice and would need to be determined experimentally to be valid. Restricting $k_{max}$ based on the expected capacitance for the monopole antenna also did not produce accurate results for other plasma conditions or the dipole case.

Therefore the issues with using the monopole antenna response function (\ref{eq:F3}) and double-sphere response function (\ref{eq:F_sph}) to predict shot noise and the antenna impedance while producing reasonable QTN predictions reveal that some physics is missing and needs to be resolved in order to use and predict these quantities more generally. The specific issue is with the $k^{-1}$ fall-off of the integrand for the real part of the impedance, which produces results for shot noise and capacitance that are non-convergent for both the monopole and double-sphere antennas. Future work must be carried out to solve the issues. One approach may be to limit $k_{max}$ on some other physical grounds. Another may be to find a more accurate representation of the current distribution for the monopole and double-sphere antennas. In addition, experimental measurement of the shot noise spectra, impedances, and response functions may clear up some of the issues. Empirical results could also be used to find an appropriate value for $k_{max}$, if it exists.

In future, we will also look at using other theoretical assumptions to derive an appropriate analytical antenna response function for monopole antennas. In particular, we plan on using the work of \citet{kellogg1981} as a starting point to derive a new function. We provide the code used to obtain the plots and results in this paper as a repository online (listed in the Acknowledgements). We hope this will prove useful for others in the community in exploring their own antenna response functions.

\section{Conclusion}
\label{sec:Conc}
The wire dipole antenna response function was re-derived in this paper following steps similar to \citet{kuehl1966} and \citet{couturier1981}. This procedure was then used to derive a new function (\ref{eq:F3}) for the monopole antenna response function, which produces reasonable predictions for the QTN spectrum.

The wire dipole response function was shown to be a well-behaved function that approached zero sufficiently rapidly ($\propto$ $k^{-2}$) for small and large values of $kL$ to produce reasonable predictions for the QTN and shot noise spectra in chosen plasma environments considered, specifically the lower ionosphere at altitudes of \SI{300}{\kilo\meter}, \SI{800}{\kilo\meter} and \SI{1500}{\kilo\meter} above the Earth's surface \citep{maj2017}. Specifically, the QTN and shot noise spectra converged as the $k$ range was increased for the integral in the impedance expression (\ref{eq:Imped_w/F1}) while the antenna capacitance that was calculated using this impedance was within 4\% of the approximation determined by \citet{meyer1989}. 

The derived monopole response function, on the other hand, approached a constant for large $kL$, very similar to the double-sphere antenna response function which was explored as a point of comparison. Both the monopole and double-sphere function produced ill-defined predictions for the shot noise and capacitance. The imaginary component of (\ref{eq:Imped_w/F}) does not converge to zero as quickly as the wire dipole function for large values of the integration variable $k$ ($\propto$ $k^{-1}$ versus $\propto$ $k^{-2}$, respectively). This produced non-convergent results for the shot noise spectrum level and capacitance - as the integration domain increases the shot noise level increases while the capacitance decreases. However, the monopole QTN spectrum does converge since this is determined by the real part of the impedance, which has a different functional form that does allow convergence at large $k$. 

Restricting the upper limit (maximum $k$) of the $k$ integrals based on physical grounds was investigated as one way to create finite results for the shot noise and capacitance. However, using the Debye length, antenna length or the capacitance as criterion for restriction does not produce acceptable results. Specifically, the wire dipole results for shot noise and capacitance are not reproduced well using a wavenumber restriction, revealing a lack of generality for these criterion. Experimental testing would be one way to find such a $k$ restriction or to verify the validity of a yet unknown general monopole response function. We provide an online repository of the code used to produce our results so that others may explore their own expressions.

\acknowledgments
R. Maj was financially supported under a University of Sydney Postgraduate Awards (SC0649) scholarship.
M. M. Martinovi\'c was financially supported by the Ministry of Education, Science and Technological Development of Republic of Serbia through financing the project ON176002 and by NASA grant 80NSSC19K0521. The code used for calculations and figures in this paper can be found at:
\url{https://github.com/ronaldmaj/Monopole_Paper_Calc_Fig_Script} - the code for calculating kappa distributions within this was adapted from \citet{odelstad2013}.
The International Reference Ionosphere (IRI) was used as a source of model data for the electron density and temperature values in this paper: \url{https://ccmc.gsfc.nasa.gov/modelweb/models/iri2016_vitmo.php}.

\listofchanges

\end{document}